\documentclass[twocolumn,groupedaddress,
superscriptaddress,preprintnumbers,
amsmath,amssymb,
aps
]{revtex4}
\usepackage{graphicx}
\usepackage{dcolumn}
\usepackage{multirow}
\usepackage{bm}%
\usepackage[colorlinks=true,citecolor=blue]{hyperref}
\hypersetup{colorlinks=true,citecolor=blue,linkcolor=red,urlcolor=blue}
\usepackage{braket}
\usepackage{textcomp}
\usepackage{times}
\usepackage{amsmath,amssymb}
\usepackage{float}
\usepackage{amsmath}

\usepackage{color}
\usepackage{color,soul}
\definecolor{Green}{RGB}{0,204,102}
\definecolor{Purple}{RGB}{102,0,255}
\definecolor{Blue}{RGB}{51,153,255}
\definecolor{Red}{RGB}{151,010,010}

\makeatletter
\def\@bibdataout@aps{%
\immediate\write\@bibdataout{%
@CONTROL{%
apsrev41Control%
\longbibliography@sw{%
    ,author="08",editor="1",pages="1",title="0",year="1"%
    }{%
    ,author="08",editor="1",pages="1",title="",year="1"%
    }%
  }%
}%
\if@filesw \immediate \write \@auxout {\string \citation {apsrev41Control}}\fi
}
\makeatother

\begin{document}

\title{Magnetic domain wall dynamics under external electric field in bilayer CrI$_3$}

\author{Sahar Izadi Vishkayi}
\affiliation{School of Physics, Institute for Research in Fundamental Sciences (IPM), Tehran 19395-5531, Iran}
\author{Reza Asgari}
\email{asgari@ipm.ir}
\affiliation{School of Physics, Institute for Research in Fundamental Sciences (IPM), Tehran 19395-5531, Iran}
\affiliation{School  of  Physics,  University  of  New  South  Wales,  Kensington,  NSW  2052,  Australia}

\begin{abstract}
Motivated by manipulating the magnetic order of bilayer CrI$_3$, we carry out microscopic calculations to find the magnetic order and various magnetic domains of the system in the presence of an electric field. Making use of density functional simulations, a spin model Hamiltonian is introduced consisting of isotropic exchange couplings, Dzyaloshinskii-Moriya (DM) interaction, and on-site magnetic anisotropy. The spin dynamics of two well-known states of bilayer CrI$_3$, low temperature (LT) and high temperature (HT) phases, are obtained by solving the Landau-Lifshitz-Gilbert equation. We show that the magnetic texture is stacking-dependent in bilayer CrI$_3$ and stable magnetic domains can appear in the HT stack which are tunable by external electric and magnetic fields. Therefore, we suggest that the HT phase represents a promising candidate for data storage in the modern generation of spintronic devices working on magnetic domain engineering.
\end{abstract}

\maketitle


\section{Introduction}\label{sec:intro}
 
Two-dimensional (2D) magnetic materials have been investigated by experimental \cite{huang2017layer, deng2018gate,gong2017discovery,gibertini2019magnetic,gong2019two,meseguer2021coexistence} and theoretical \cite{Hxxz, Morell2019, torelli2020first,frey2019surface, menichetti2019electronic} researchers to analyze their attractive properties from the initial report of the successful synthesis of thin layer CrI$_3$ \cite{huang2017layer}.
The presence of a substantial magnetocrystalline anisotropy was discovered, thus breaking the invariance under spin rotations, and the crystal magnetic ordering was observed to be a layer-dependent phenomenon.
Because of the existence of metastable magnetic domains with ultra-thin domain walls (like permanent nanoparticle magnets \cite{hubert2008magnetic}) which possess spin dynamics \cite{wahab2021quantum,PhysRevX.11.031047,akram2021moire,xu2021emergence}, CrI$ _3$, in monolayer and multilayer forms, has recently been described as a viable contender for the modern generation of memory devices.
The microscopic nature of these magnetic domains in the presence of external fields is still poorly understood, offering a significant challenge for theoretical materials science. 

Two geometrically stable stacks of bilayer CrI$_3$ structures are known as low-temperature (LT) and high-temperature (HT) phases (in more than $210$ K) \cite{Morell2019, jiang2019stacking,kim2019evolution}. The LT phase is a ferromagnetic (FM) ground state in which the magnetic moments of the Cr atoms are aligned in both layers, whereas the magnetic moments of the Cr atoms in distinct layers of the HT phase are in an antiferromagnetic (AFM) configuration \cite{Morell2019, lei2021magnetoelectric, song2019switching, li2019pressure, sivadas2018stacking}.
The competition between orbitally dependent interlayer AFM super-super-exchange and interlayer FM super-super-exchange causes the magnetic ordering. In the HT bilayer CrI$_3$, a vertical external electric field generates an AFM to an FM magnetic transition \cite{Jiang2018mater, Morell2019,jiang2018controlling}. Because of their sensitivity to the electric field, the HT phase of the bilayer CrI$_3$ can be more efficient in ultrafast memory devices, and it could be the first device for the engineering of magnetic domains in the world of 2D magnets.
For a monolayer up to bulk structures, the Curie temperature of a thin layer CrI$_3$ ranges from $45$ K to $61$ K \cite{huang2017layer, kashin2020orbitally}. The magnetic ordering temperature of a thin layer CrI$_3$ is raised by the number of layers, according to experiments \cite{huang2017layer}. The N{\'e}el temperature of $45$ K is reported for the HT bilayer \cite{huang2018electrical} and it is expected that the LT bilayer maintains a larger Curie temperature due to strong FM ordering between the layers \cite{wang2021systematic}.

Although the Ising Hamiltonian model is a classic model for explaining the magnetic behavior of 2D CrI$_3$, it cannot explain the experimentally observed spin wave \cite{LebingChen}. To explain the spin canting physics, an extended Heisenberg Hamiltonian containing Kitaev \cite{lee2020fundamental}, biquadratic anisotropy exchange term \cite{wahab2021quantum}, or Dzyaloshinskii-Moria (DM) \cite{vishkayi2020strain, jaeschke2021theory} interaction terms was proposed.
The spin wave in a CrI$_3$ structure \cite{chen2020magnetic} is described by a Heisenberg Hamiltonian with Kitaev or DM interactions, and a small portion of the DM interactions in 2D CrI$_3$ is responsible for a topological gap \cite{jaeschke2021theory}. 

The ability to alter and adjust magnetic domains of 2D bilayer magnets to make them useful in technology and spintronics is a key question here.
We focus on the bilayer CrI$_3$ in the presence of an external electric field \cite{huang2018electrical, Jiang2018mater,jiang2018controlling, Morell2019, lei2021magnetoelectric} to answer this specific topic. 

In this study, we use a spin model Hamiltonian in the presence of external fields, with isotropic exchange and DM interactions.
We use density functional theory (DFT) computations to derive a suitable spin model Hamiltonian for the bilayer CrI$_3$ that accurately reproduces the magnetic characteristics of the system.
The isotropic exchange coupling parameters of the bilayer CrI$_3$ have been investigated in the presence of an external electric field \cite{lei2021magnetoelectric}, but the Dzyaloshinskii-Moria (DM) interaction has yet to be investigated.
In contrast to what we found for a monolayer CrI$_3$ \cite{vishkayi2020strain}, our calculations suggest that the DM interaction in the HT bilayer CrI$_3$ is substantially connected to the applied electric field. The spin dynamics of the system are calculated using the Landau-Lifshitz-Gilbert formalism \cite{ellis2015landau} with DFT-derived exchange parameters as starting inputs. To obtain a temperature dependent magnetization, we also employ the Monte Carlo Metropolis technique with $10^6$ time steps for a $50\times 50$ nm$^2$ layer.

 The weak strength of the isotropic intralayer exchange coupling signifies the magnetic ordering of the layers, which can be modified by an external electrical field, according to our findings. The findings also reveal that an electric field, like a magnetic field, alters the magnetic texture of the system.
Furthermore, by applying an electric field resulting from an increase in the intralayer isotropic exchange coefficient, the temperature of the magnetic order of the HT bilayer is raised. The electric field also affects the magnetic texturing on the domain walls. The results reveal that the magnetic textures in the studied systems can be manipulated using an external electric field. To be more precise, in the LT bilayer, we show that the DM interactions lead to the creation of a quasi-circular domain shape in the layers, but its value is extremely smaller than the intralayer exchange coupling to stabilize it in the system, therefore, the obtained magnetic domain is unstable and disappearing quickly.

\begin{figure}
	\begin{center}
		\includegraphics[width=8.5cm]{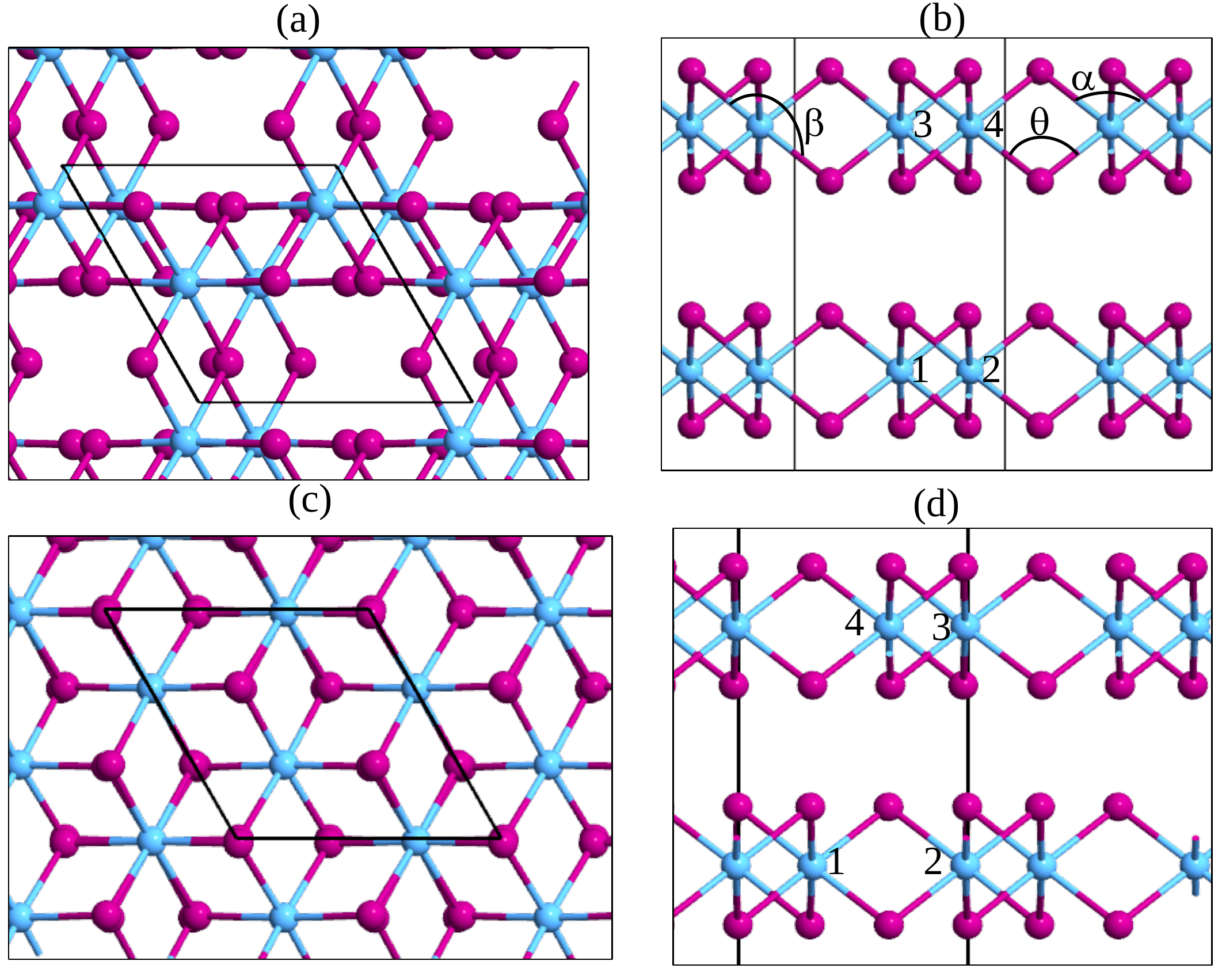}
		\caption{(Color online) The top view ((a) and (c)) and side view ((b) and (d)) of the HT and LT bilayer CrI$_3$ structures, respectively. The black lozenges show the unit cell of bilayers. The blue (pink) circles represent chromium (iodine) atoms. In (b) and (d), chromium atoms are labeled by 1-4 in the unit cell. In (b) the bonding angles between the atoms of each layer are shown by $\alpha$, $\beta$ and $\theta$. The crystal parameters are given in Table I.}
		\label{geometry}
	\end{center}
\end{figure}

This paper is organized as follows. We commence with a description of our theoretical formalism in Sec. II, followed by the details of the DFT simulations, the spin model Hamiltonian and spin dynamics. Numerical results for the spin-spin interaction parameters in the presence of an electric field, the exchange coefficients in both the HT and LT phases and spin dynamics in two cases are reported in Sec. III. We summarize our main findings in Sec. IV.

\section{Theory and Models}
\subsection{DFT calculations}
We use Quantum Espresso package \cite{refQE} to simulate density-functional based calculations. The generalized gradient exchange-correlation within Perdew-Burke-Ernzerhof \cite{refpbe} functional is considered to expose the electronic and magnetic ground states of two bilayer CrI$_3$ structures. An energy cut-off of $50$ Ry for wave vector and an $8\times8\times1$ grid mesh of $k$-points within the first Brillouin zone are defined as the converged input parameters. To avoid repeats effects, we consider a vacuum of $20$ \AA~ along the $z$-direction. The van der Waals Grimme-D2 \cite{grimme2006semiempirical} correction is used to consider the interaction between layers. DFT$+U$ calculations \cite{cococcioni2005linear} are also needed to find correct magnetic ground states of CrI$_3$ bilayers. At this point, we consider $U=3$ eV as the on-site Hubbard parameter similar to Refs. \cite {jiang2019stacking, lei2021magnetoelectric, leon2020strain, akram2021moire}. To obtain a reliable total ground-state energy, we maintain a high degree of accuracy of $10^{-10}$  eV. The bilayers are relaxed until the maximum force on each atom is $0.01$  eV/\AA ~  even under the electric field conditions. It should be noted that the consideration of the spin-orbit coupling (SOC) and noncollinear spin-polarization are essential to obtain the DM interaction and magnetic anisotropic energy (MAE).

\subsection{Spin-model Hamiltonian}
A spin model Hamiltonian for a 2D magnetic hexagonal lattice in each layer is utilized to provide spin-spin interactions in a magnetic 2D bilayer CrI$_3$ system:
\begin{align}
{\cal H}= &\sum_{i,j} [ \frac{1}{2}J_{ij}{\bf S}_i\cdot{\bf S}_j+ {\gamma}_{i}|S_{iz}|^2+ \nonumber\\
& \frac{1}{2}{\bf D}_{ij}\cdot({\bf S}_i\times{\bf S}_j )+\sum_i \mu_B{{\bf B}_{ext}}\cdot{\bf{S}}_i],
\label{eq1}
\end{align}
where $J_{ij}$ is the isotropic exchange coupling parameter between $i$ and $j$ atoms, $\gamma_i$ denotes a magnetic anisotropy coefficient, ${\bf D}_{ij}$ is the DM interaction vector and the last term shows the effect of an external magnetic field, ${\bf B}_{ext}$, and $\mu_B$ is the magnetic Bohr. The magnetic moment of i$^{th}$ Cr atom is indicated by ${\bf S}_i$ in Eq. (\ref{eq1}). It is intriguing to compare the strength of the interlayer and intralayer exchange coefficient parameters which are respectively referred by $v$ and $t$ indices in this paper.

\subsection{Spin Dynamics}
The Landau-Lifshitz-Gilbert (LLG) equation \cite{ellis2015landau} is used to atomistic model the spin dynamics of the system after the coefficients of the spin Hamiltonian are established. We follow the method introduced by Evans et al. \cite{evans2014atomistic} implemented in VAMPIRE code. The time evolution of the spin direction is given by LLG equation according to
\begin{equation}
\frac{\partial{\bf S}_i}{\partial t}=-\frac{\gamma}{1+\lambda^2}[{\bf S}_i\times{\bf H}_{eff}^i+\lambda {\bf S}_i \times ({\bf S}_i \times {\bf H}_{eff}^i)],
\label{eq2}
\end{equation}
where $\gamma$ is the gyromagnetic ration and $\lambda$ is microscopic damping equal to unity for finding the equilibration states quickly.
The effective field applied on each spin is defined by ${\bf H}_{eff}^i=-{\partial \cal H}/{\partial {\bf S}_i}+{\bf H}_{th}^i$ and the effective thermal field, ${\bf H}_{th}^i$, is counted by Langevin dynamic approach \cite{brown1979thermal} to consider thermal spin fluctuations.

In order to obtain the spin dynamics simulations, a 100 nm $\times$ 100 nm square bilayer CrI$_3$ was kept in thermal equilibrium above the critical (N{\'e}el or  Curie) temperature and then linearly cooled to 0 K in 2 ns. The magnetic moments of atoms are randomized at a temperature above the critical temperature, so we chose random initial spin directions for the spin dynamics simulation. The time step of 1 fs  is considered in the integration scheme of Heun's method \cite{garcia1998langevin} for solving LLG equations, suitable for materials with low critical temperature and large dimensions to obtain stable results \cite{evans2014atomistic}.
	
 The long-range demagnetization field is also considered in the spin dynamics calculations. The magnetic dipole-dipole interaction induces local spin-flip in the magnetic materials and leading to the creation of complicated spin texture even in the materials with a FM ground state \cite{kawaguchi2007can, ezawa2010giant}. We discretize the layer to macrocells with $1$ nm$^3$ volume to avoid expenses of the demagnetization field computing at the atomistic level. The magnetic moments of macrocells, ${\bf m}_{mc}$, are defined by the sum of atomic magnetic moments within the macrocell and the demagnetization field of i$^{th}$ macrocell is given by
\begin{equation}
{\bf H}_d^{mci}=
{\frac {{\mu}_0}{4 \pi}} (\sum_{i\neq j}{\bf M}_{ij}\cdot{\bf m}_{mc}^j)-{\frac {\mu_0} {3} }
{\frac {{\bf m}_{mc}^i}{V_{mc}^i}},
\label{eq3}
\end{equation}
where $V_{mc}^i$ is the volume of i$^{th}$ macrocell, $\mu_0=4\pi\times10^{-7}$ is the vacuum permeability  and ${\bf M}_{ij}$ is the dipole-dipole interaction matrix between $i$ and $j$ macrocells defined as \cite{evans2014atomistic},
\begin{equation}
{\bf M}_{ij}=
\begin{bmatrix}
\frac{3r_xr_x-1}{r_{ij}^3}-\frac{1}{3} & 3 r_x r_y & 3 r_x r_z \\
3 r_x r_y & \frac{3r_yr_y-1}{r_{ij}^3}-\frac{1}{3} & 3 r_y r_z \\
3 r_x r_z & 3 r_y r_z &\frac{3r_zr_z-1}{r_{ij}^3}-\frac{1}{3}
\end{bmatrix},
\end{equation}
where the positions of macrocells are obtained by the magnetic center of the mass relation and the vector between the position of $i$ and $j$ macrocells is shown by ${\bf r}_{ij}=r_{ij} (r_x \hat{i}+r_y \hat{j}+r_z \hat{k}) $.
We consider that the sheet is linearly cooled in $2$ ns from an equilibrium temperature, upper the Curie temperature, to zero Kelvin. To obtain temperature dependent magnetization, we use the Monte Carlo Metropolis algorithm with $10^6$ time steps for a $50$$\times 50$ nm$^2$ sheet.

\section{Numerical Results and Discussions}
The magnetic ground state, symmetric and asymmetric exchange coefficients, and temperature-dependent magnetization of bilayer CrI$_3$ in both the HT and LT phases in the ground state are presented in this section.
The effects of an external electric field on the coefficients are then studied.
Finally, we investigate the spin dynamics of both phases in the presence of a perpendicular electric field.

The HT phase of bilayer CrI$_3$ is in the form of the monoclinic structure, while the LT bilayer possesses a rhombohedral crystal structure as displayed in Figs. \ref{geometry}(a)-(d). The different crystal structures leads to the unique magnetic behaviors. Both the HT and LT phases are relaxed by spin-dependent DFT calculations and their optimized lattice constant, bonding lengths and angles are reported in Table \ref{tabgeo}. The results reveal that the bonding angles and lengths of both layers are the same, and that the geometry of the bilayer is unaffected by the electric field, which is in good agreement with those described in Ref. \cite{Morell2019}. To obtain a formation of the energy, we use the straight formula of $E_{\rm F}=E_t(bilayer)-n_{Cr}E(Cr)-n_{I}E(I)$ where $E_t(bilayer)$ is the total energy of the bilayer CrI$_3$, $n_{Cr} (n_{I})$ is the number of the chromium (iodine) atoms in the unit cell, and $E(Cr)$ and $E(I)$ are the energy of free Cr and I atoms, respectively. The formation energies of the HT and LT bilayers are given in Table \ref{tabgeo} indicating that the both phases are energetically favored to occur.
\begin{table*}
	\caption{Lattice constant, $a$, Cr-Cr (I-I) distance between the layers, $h_{Cr-Cr} (h_{I-I})$, Cr-I (Cr-Cr) bonding length, $d_{Cr-I} (d_{Cr-Cr})$, 
	 two different I-Cr-I bonding angles, $\beta_{I-Cr-I}$ and $\alpha_{I-Cr-I}$, and Cr-I-Cr bonding angle, $\theta_{Cr-I-Cr}$, and formation energy, $E_f$ of the HT and LT bilayer CrI$_3$ structures. The bonding angles are shown in Fig. \ref{geometry} (b).  }
	\begin{center}
		\begin{tabular}{|c|c|c|c|c|c|c|c|c|c|}
			\hline
			\hline
			~~Phase~~ & ~~ $a (\AA)$ ~~ & $h_{Cr-Cr} (\AA)$ & $h_{I-I} (\AA)$ & $d_{Cr-I} (\AA)$ & $d_{Cr-Cr}(\AA)$  & $\beta_{I-Cr-I} (^o)$ & $\alpha_{I-Cr-I} (^o)$ &
			  $\theta_{Cr-I-Cr} (^o)$ & $E_F$ (eV/per Cr) \\
			\hline
			HT & 6.977 & 7.24 & 3.99 & 2.81 & 4.03 & 177.4 & 89.9 & 91.6 & -8.81 \\
			\hline
			LT & 6.979 & 6.59 & 3.35 & 2.81 & 4.03 & 177.6 & 89.9 & 91.7 & -8.86 \\
			\hline
			\hline
		\end{tabular}
	\end{center}
	\label{tabgeo}
\end{table*}
\subsection{Exchange coefficients in the HT phase}
\begin{figure}
	\begin{center}
		\includegraphics[width=8.5cm]{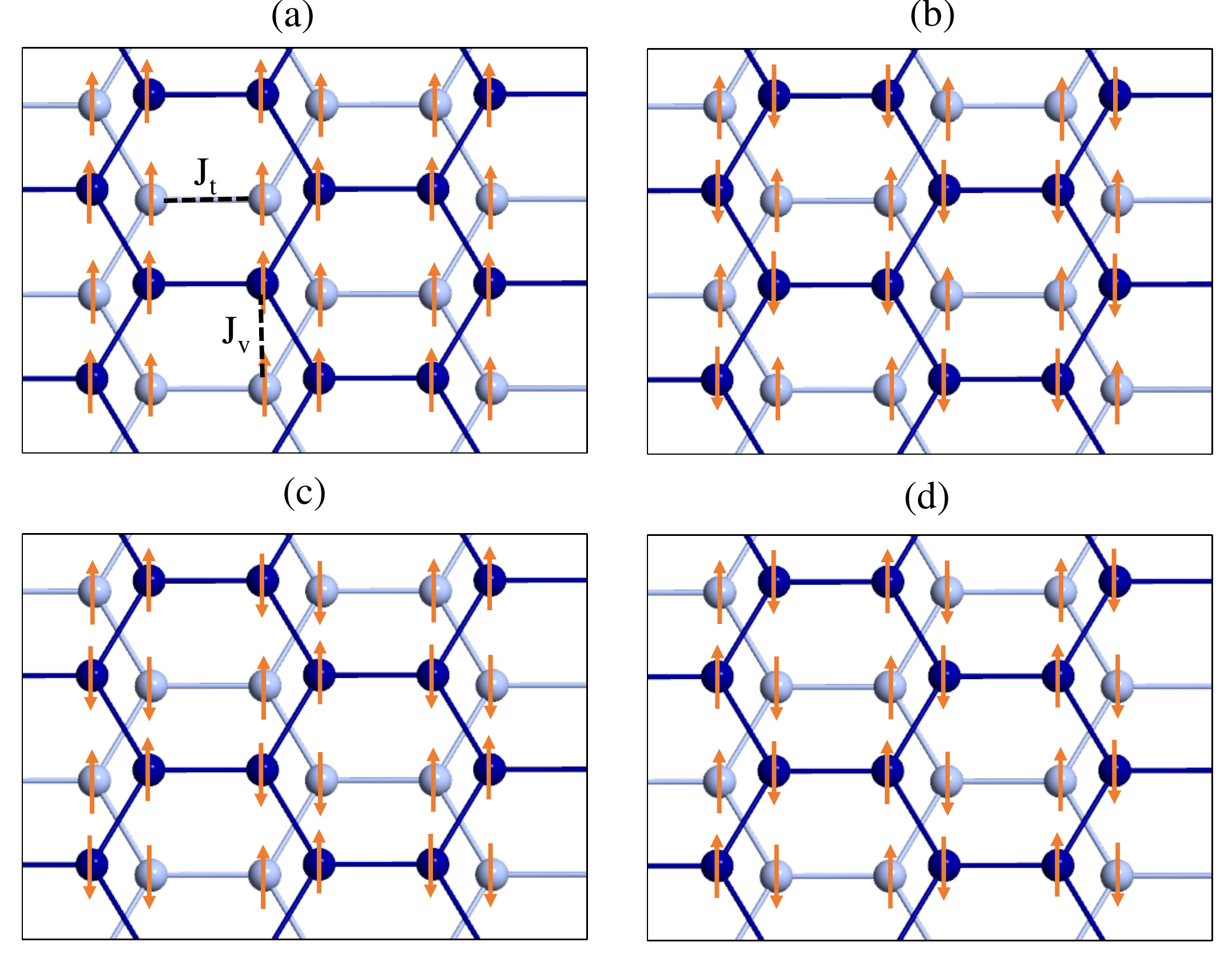}
		\caption{(Color online) The magnetic ordering of Cr atoms in (a) FM, (b) AFM1, (c) AFM2 and (d) AFM3 spin configurations of HT bilayer CrI$_3$. The dark (light) balls represent the Cr atoms in the top (bottom) layer and arrows show the magnetic moment vectors of Cr atoms in the z-direction. The dashed lines connect the atoms with intralayer and interlayer exchange coupling which are indicated, respectively, by t and v indices.}
		\label{magneticorder}
	\end{center}
\end{figure}

In the HT phase of bilayer CrI$_3$, there are four Cr atoms in the unit cell; two in each layer.
At zero external magnetic field, if the nearest neighbor interlayer and intralayer interactions are considered, the extended first two terms of the model Hamiltonian (Eq. (\ref{eq1})) in the unit cell can be written as
\begin{align}
	{\cal H}=& \frac{1}{2}(\eta_t J_t [({\bf S}_1\cdot{\bf S}_2)+({\bf S}_2\cdot{\bf S}_1)+({\bf S}_3\cdot{\bf S}_4)+({\bf S}_4\cdot{\bf S}_3)]+ \nonumber \\
	&\eta_{1v} J_{1v} 
	[({\bf S}_1 \cdot {\bf S}_3)+({\bf S}_3 \cdot {\bf S}_1)+({\bf S}_2 \cdot {\bf S}_4)+({\bf S}_4 \cdot {\bf S}_2)]+ \nonumber \\
	&\eta_{2v} J_{2v} 
	[({\bf S}_1 \cdot {\bf S}_4)+({\bf S}_4 \cdot {\bf S}_1)+({\bf S}_2 \cdot {\bf S}_3)+({\bf S}_3 \cdot {\bf S}_2)])+ \nonumber \\
	& \eta \gamma |S_z|^2	
\end{align}
where $\eta_t=3$ is the number of the nearest neighbors in each layer and $\eta_{1v}=\eta_{2v}=1$ are the number of type-$1$ (between atoms numbered by $1$ ($2$) and $3$ ($4$) in Fig. \ref{geometry} (b)) and type-$2$ (between atoms numbered by $1$ ($2$) and $4$ ($3$) in Fig. \ref{geometry} (b)) neighbors in the interlayer. $\eta=4$ is the number of Cr atoms in the unit cell.
The exchange coupling coefficients are obtained by mapping the model Hamiltonian onto the DFT-based calculations which are computed for various magnetic configurations.
For calculation of the isotropic exchange coefficient, $4$-distinct spin configurations are considered as;
 FM (${\bf S}_1=S \hat{k}, {\bf S}_2=S \hat{k}, {\bf S}_3=S \hat{k}, {\bf S}_4=S \hat{k}$), 
AFM1 (${\bf S}_1=S \hat{k}, {\bf S}_2=S \hat{k}, {\bf S}_3=-S \hat{k}, {\bf S}_4=-S \hat{k}$), 
AFM2 (${\bf S}_1=S \hat{k}, {\bf S}_2=-S \hat{k}, {\bf S}_3=-S \hat{k}, {\bf S}_4=S \hat{k}$) and
AFM3 (${\bf S}_1=S \hat{k}, {\bf S}_2=-S \hat{k}, {\bf S}_3=S \hat{k}, {\bf S}_4=-S \hat{k}$) in which all the magnetic moments of Cr atoms are aligned in the $z$-direction and $S=3/2$. In Fig. \ref{magneticorder}, the considered magnetic orders are represented in bilayer CrI$_3$.
In these configurations, the DM terms of Eq. (\ref{eq1}) are vanished owing to the parallel magnetic moments of Cr atoms, and their total energies are obtained by
\begin{align}
&E_{FM}=2S^2(\eta_t J_t+\eta_{1v}J_{1v}+\eta_{2v}J_{2v})+\eta\gamma S^2, \\
&E_{AFM1}=2S^2(\eta_t J_t-\eta_{1v}J_{1v}-\eta_{2v}J_{2v})+\eta\gamma S^2, \nonumber \\
&E_{AFM2}=2S^2(-\eta_t J_t-\eta_{1v}J_{1v}+\eta_{2v}J_{2v})+\eta\gamma S^2, \nonumber \\
&E_{FM3}=2S^2(-\eta_t J_t+\eta_{1v}J_{1v}-\eta_{2v}J_{2v})+\eta\gamma S^2.\nonumber 
\end{align}
We carry out spin-dependent DFT calculations to find the total energies of the considered magnetic configurations and as a consequence the exchange coupling parameters of the HT bilayer are given by
\begin{align}
&J_t=\frac{(E_{FM}+E_{AFM1})-(E_{AFM2}+E_{AFM3})}{8S^2\eta_t}, \\
&J_v=\frac{E_{FM}-E_{AFM1}}{4S^2\eta_v},
\label{Jt_JvHT}
\end{align}
where $J_v=J_{1v}+J_{2v}$ is the interlayer exchange coupling and $\eta_v=\eta_{1v}=\eta_{2v}$.
The intralayer exchange coupling is $J_t=-7.00$ meV which presents the strong ferromagnetic coupling of Cr atoms in the layers.
$J_v=+0.08$ meV indicates the layers are coupled in a form of antiferromagnetic.
Furthermore, the weak strength of the isotropic interlayer exchange coupling denotes the magnetic order of the layers can be tuned by the external electrical field \cite{Morell2019}.

\begin{figure}
	\begin{center}
		\includegraphics[width=8.5cm]{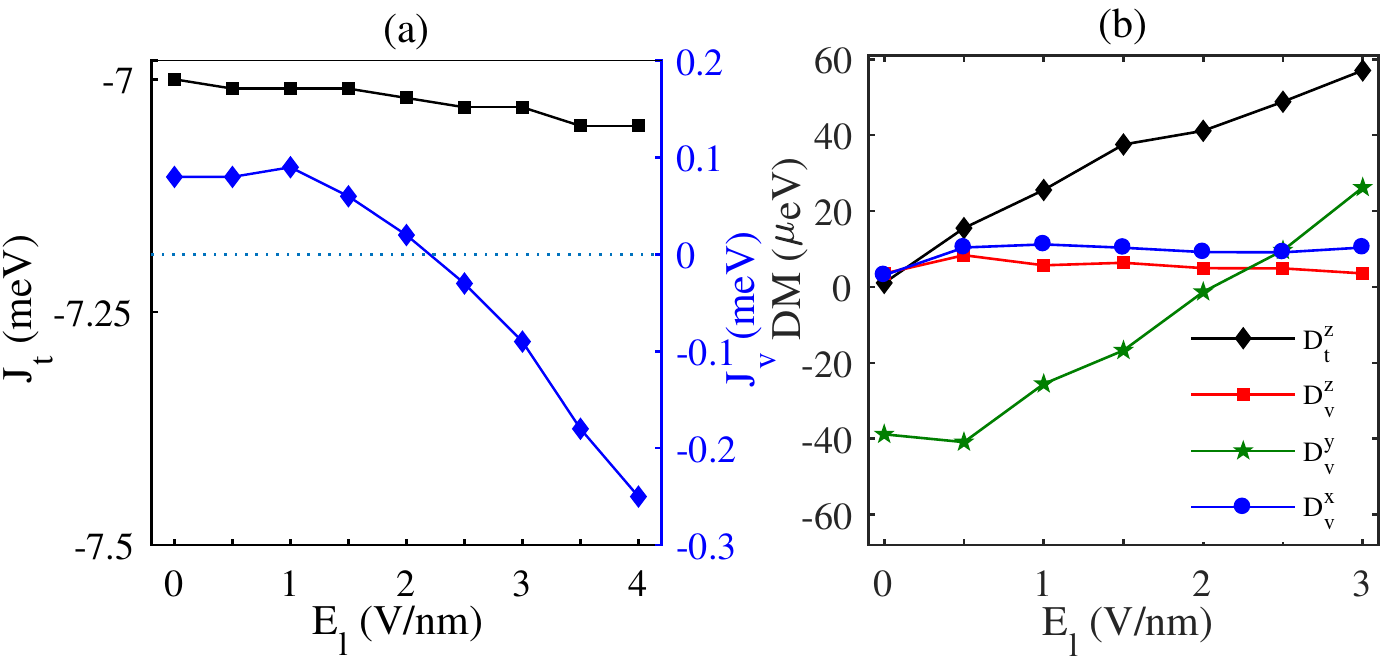}
		\caption{(Color  online) (a) The intralayer (squares) and interlayer (lozenges) isotropic exchange coefficients as a function of the external electric field, $E_l$. 
		The change of the sign of the interlayer exchange shows a magnetic transition from the AFM to a FM between the layers. The ferromagnetic coupling of Cr atoms in the layer is extremely strong. (b) The non-zero component of interlayer and intralayer DM vectors of the bilayer CrI$_3$ structures as a function of $E_l$. The intralayer exchange coupling is changed significantly by the electric field. Although the $z$ and $x$ components of the interlayer DM interactions are independent of the electric field, the sign and the value of $D_v^y$ vary.  }
		\label{J_DMHT}
	\end{center}
\end{figure}

We observe the variation of the isotropic interlayer and intralayer exchange coupling by DFT-based calculations, shown in Fig. \ref{J_DMHT}.
It is observed that the sign of the interlayer exchange coupling is changed between $2$-$2.5$ V/nm, hence a magnetic transition from the AFM to the FM occurs between layers while the strength of the intralayer exchange coefficient is no longer altered by the electric field.
This means that the ferromagnetic coupling of the Cr atoms in the layer is extremely strong.

The extended Hamiltonian for the DM term is given by
\begin{align}
	{\cal H}_{DM}=&\frac{1}{2}[{\bf D}_{t12}\cdot({\bf S}_1\times{\bf S}_2)+{\bf D}_{t12'}\cdot({\bf S}_1\times{\bf S}_{2'})+{\bf D}_{t12''}\cdot \nonumber \\
	&({\bf S}_1\times{\bf S}_{2''})+
	{\bf D}_{t21}\cdot({\bf S}_2\times{\bf S}_1)+{\bf D}_{t21'}\cdot({\bf S}_2\times{\bf S}_{1'}) \nonumber \\
	&+{\bf D}_{t21''}\cdot({\bf S}_2\times{\bf S}_{1''})+
	{\bf D}_{t34}\cdot({\bf S}_3\times{\bf S}_4)+{\bf D}_{t34'}\cdot \nonumber \\
	&({\bf S}_3\times{\bf S}_{4'}) 
	+{\bf D}_{t34''}\cdot({\bf S}_3\times{\bf S}_{4''})+
	{\bf D}_{t43}\cdot({\bf S}_4\times{\bf S}_3) \nonumber \\
	&+{\bf D}_{t43'}\cdot({\bf S}_4\times{\bf S}_{3'}) 
	+{\bf D}_{t43''}\cdot({\bf S}_4\times{\bf S}_{3''})+  \nonumber \\
	&{\bf D}_{v13}\cdot({\bf S}_1\times{\bf S}_{3})+ {\bf D}_{v31}\cdot({\bf S}_3\times{\bf S}_{1})+ \nonumber \\
	&{\bf D}_{v14}\cdot({\bf S}_1\times{\bf S}_{4})+ {\bf D}_{v41}\cdot({\bf S}_4\times{\bf S}_{1})+ \nonumber \\
    &{\bf D}_{v23}\cdot({\bf S}_2\times{\bf S}_{3})+ {\bf D}_{v32}\cdot({\bf S}_3\times{\bf S}_{2})+ \nonumber \\
    &{\bf D}_{v24}\cdot({\bf S}_2\times{\bf S}_{4})+ {\bf D}_{v42}\cdot({\bf S}_4\times{\bf S}_{2})],
    \label{HDMHT}
\end{align}
where ${\bf S}_{i'}$ and ${\bf S}_{i''}$ are equal to the magnetic moment of i$^{th}$ Cr atom, and ${\bf S}_{i'}={\bf S}_{i''}={\bf S}_{i}$.
On the other hand, the DM vector between two atoms can be obtained by ${\bf D}_{ij}=D_{ij} {\hat{\bf u}}_{ij}$, where $D_{ij}=D_{ji}$ is a scalar and ${\hat{\bf u}}_{ij}$ is a unit-vector in the direction of the line connecting $i$ and $j$ atoms \cite{yang2015anatomy}.
Therefore, ${\bf D}_{ij}=-{\bf D}_{ji}$, ${\bf D}_{t12'}={\bf R}_z(120) {\bf D}_{t12}$, ${\bf D}_{t12''}={\bf R}_z(240) {\bf D}_{t12}$ and ${\bf R}_z(\theta)$ is a $\theta$ degree rotation around the $z$-axis. Three first terms of Eq. (\ref{HDMHT}) can be written as
\begin{align}
{\cal H}&_{DM12}= \\ \nonumber
&({\bf D}_{t12}+{\bf R}_z(120){\bf D}_{t12}+{\bf R}_z(240){\bf D}_{t12})\cdot({\bf S}_1\times{\bf S}_2).
\end{align}
If ${\bf D}_{t12}=(D_{t12}^x \hat{i} +D_{t12}^y \hat{j} +D_{t12}^z \hat{k})$, then $({\bf D}_{t12}+{\bf R}_z(120) {\bf D}_{t12}+ {\bf R}_z(240) {\bf D}_{t12})=3D_{t12}^z \hat{k}$, hence only the $z$-component of the DM vector between two Cr atoms in the layers of the bilayer CrI$_3$ can be entered in the model Hamiltonian of Eq. (\ref{eq1}). In the similar way, we can write $({\bf D}_{t34}+{\bf R}_z(120){\bf D}_{t34}+{\bf R}_z(240){\bf D}_{t34})=3D_{t34}^z \hat{k}$  for the DM interaction between Cr atoms in the top layer. Due to the symmetries between the layers, it is possible to consider $D_{t12}^z=D_{t34}^z=D_t^z$. On the other hand, ${\bf D}_{t12}=-{\bf D}_{t21}$ and ${\bf D}_{t13}=-{\bf D}_{t31}$, hence for the first five lines of Eq. (\ref{HDMHT}), we have the intralayer terms of the DM interaction as,
\begin{equation}
	{\cal H}_{DMt}=\frac{1}{2}[6D_t^z \hat{k}\cdot ({\bf S}_1\times{\bf S}_2)+6D_t^z \hat{k}\cdot ({\bf S}_3\times{\bf S}_4)].
	\label{DMtHT}
\end{equation}
Therefore, the symmetries of the HT phase dictate that just the $z$-component of the intralayer DM interaction vector, $D_t^z$, has a non-zero value.

\begin{figure}
	\begin{center}
		\includegraphics[width=8.5cm]{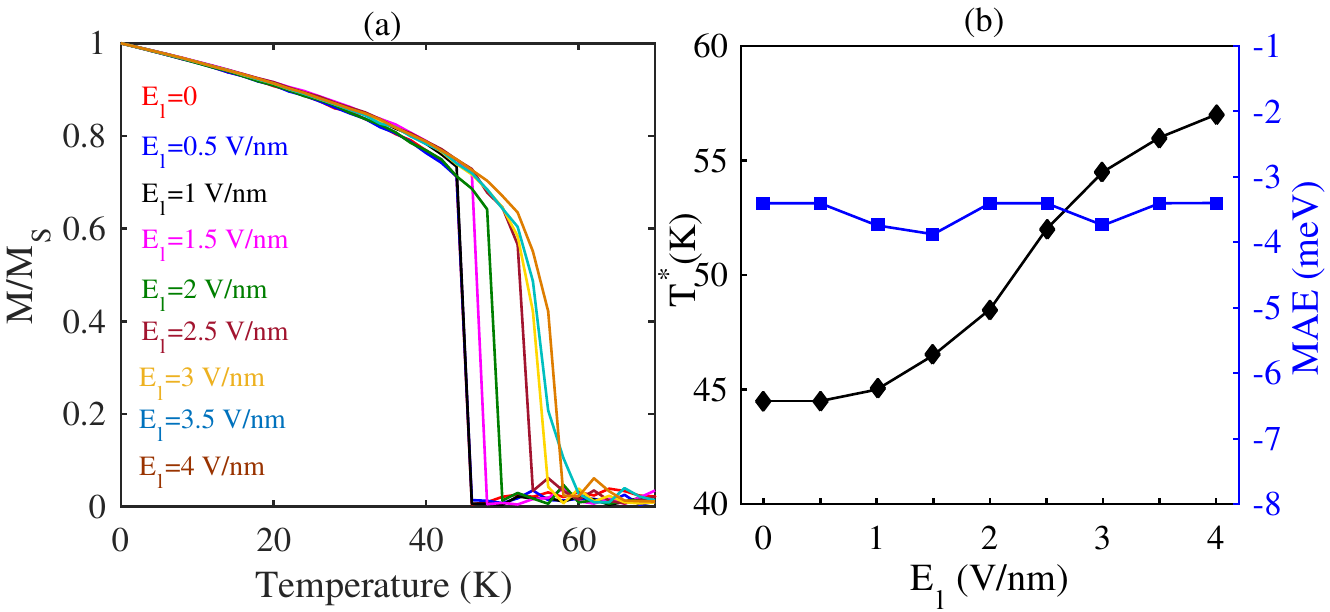}
		\caption{(Color  online) (a) The normalized magnetization of the HT bilayers as a function of temperature under various external electric field. (b) The magnetic order temperature, $T^*$
			 ($T^*=T_N$ for $E_l\leqslant 2$ and $T^*=T_C$ for $E_l\geqslant 2.5$), (lozenges) and magnetic anisotropy energy (squares) of the HT bilayer CrI$_3$ versus the external electric field. The magnetic order temperature of the HT bilayer is increased by applying the electric field that arises from the increasing of the intralayer isotropic exchange coefficient.
Furthermore, the magnetic moments of the Cr atoms are in out-of-plane direction and it is unchangeable by the external electric field.}
		\label{TcHT}
	\end{center}
\end{figure}
Moreover, the geometry of the bilayer crystal indicates that $\hat{\bf u}_{13}=\hat{\bf u}_{24}$ and $\hat{\bf u}_{32}={\bf R}_y(180)\hat{\bf u}_{14}$, then we can assume ${\bf D}_{v13}={\bf D}_{v24}$ and ${\bf D}_{v32}={\bf R}_y(180){\bf D}_{v14}$. To obtain the interlayer contribution of the DM vector, we have to focus on the last four lines of Eq. (\ref{HDMHT}) which can be summarized in the following form,
\begin{align}
	{\cal H}_{DMv}&={\bf D}_{v13}\cdot({\bf S}_1\times{\bf S}_3)+  
	{\bf D}_{v13}\cdot({\bf S}_2\times{\bf S}_4)+
	{\bf D}_{v14}\cdot \nonumber \\
	&({\bf S}_1\times{\bf S}_4)+
	{\bf R}_y(180){\bf D}_{v14}\cdot({\bf S}_3\times{\bf S}_2)
	\label{HDMvHT}
\end{align}
where ${\bf D}_{v13}=(D_{v1}^x, D_{v1}^y, D_{v1}^z)$,  ${\bf D}_{v14}=(D_{v2}^x, D_{v2}^y, D_{v2}^z)$ and ${\bf R}_y(180){\bf D}_{v14}=(-D_{v2}^x, D_{v2}^y, -D_{v2}^z)$. Eight different spin configurations are needed to obtain enough information about the interlayer and intralayer DM vectors given in Table \ref{TabSCHT}.
\begin{table*}
	\caption{Eight spin configurations for calculation of the interlayer and intralayer DM vectors of the HT bilayer. }
	\begin{center}
		\begin{tabular}{|c|c|c|c|c|c|c|c|c|}
			\hline
			\hline
			Config. &  $1^{st}$ & $2^{nd}$ & $3^{rd}$ & $4^{th}$ & $5^{th}$  & $6^{th}$ & $7^{th}$ & $8^{th}$ \\
			\hline
			${\bf S}_1$ & $S\hat{i}$ & $S\hat{i}$ & $S\hat{i}$ & $S\hat{i}$ & $S\hat{k}$ & $S\hat{k}$ & $S\hat{k}$ & $S\hat{k}$ \\
			\hline
		    \multirow{2}{*}{${\bf S}_2$} & $S(\cos(\frac{\pi}{6})\hat{i}+$    & $S(\cos(\frac{\pi}{6})\hat{i}-$& $S\hat{i}$ & $S\hat{i}$ & $S\hat{k}$ & $S\hat{k}$ & $S\hat{k}$ & $S\hat{k}$ \\
		     & $\cos(\frac{\pi}{6})\hat{j})$  & $\sin(\frac{\pi}{6})\hat{j})$ &   &      &     &      &          &  \\
			\hline
			\multirow{2}{*}{${\bf S}_3$} & $S\hat{i}$ & $S\hat{i}$ & $S(\cos(\frac{\pi}{6})\hat{i}+$ & $S(\cos(\frac{\pi}{6})\hat{i}-$ & $S(\cos(\frac{\pi}{6})\hat{k}+$ & $S(\cos(\frac{\pi}{6})\hat{k}-$ & $S(\cos(\frac{\pi}{6})\hat{k}+$ & $S(\cos(\frac{\pi}{6})\hat{k}-$ \\
			&   &   & $\sin(\frac{\pi}{6})\hat{j})$ & $\sin(\frac{\pi}{6})\hat{j})$ & $\sin(\frac{\pi}{6})\hat{i})$ & $\sin(\frac{\pi}{6})\hat{i})$ & $\sin(\frac{\pi}{6})\hat{j})$
			& $\sin(\frac{\pi}{6})\hat{j})$ \\
			\hline
			\multirow{2}{*}{${\bf S}_4$} & $S(\cos(\frac{\pi}{6})\hat{i}+$ & $S(\cos(\frac{\pi}{6})\hat{i}-$ & $S(\cos(\frac{\pi}{6})\hat{i}+$ & $S(\cos(\frac{\pi}{6})\hat{i}-$ & $S(\cos(\frac{\pi}{6})\hat{k}+$ & $S(\cos(\frac{\pi}{6})\hat{k}-$ & $S(\cos(\frac{\pi}{6})\hat{k}+$ & $S(\cos(\frac{\pi}{6})\hat{k}-$ \\
			&$\sin(\frac{\pi}{6})\hat{j})$    & $\sin(\frac{\pi}{6})\hat{j})$   & $\sin(\frac{\pi}{6})\hat{j})$ & $\sin(\frac{\pi}{6})\hat{j})$ & $\sin(\frac{\pi}{6})\hat{i})$ & $\sin(\frac{\pi}{6})\hat{i})$ & $\sin(\frac{\pi}{6})\hat{j})$
			& $\sin(\frac{\pi}{6})\hat{j})$ \\
			\hline
			\hline
		\end{tabular}
	\end{center}
	\label{TabSCHT}
\end{table*}

The DM interactions are calculated by mapping the considered spin configurations on the model Hamiltonian and comparing the spin-dependent DFT-obtained total energies. At this place, we want to focus on the DM terms described in Eqs. (\ref{DMtHT}) and (\ref{HDMvHT}).
\begin{align}
	& {\cal H}_{DMt(1^{st})}=6S[(D_t^z \hat{k})\cdot (\sin(\pi/6) \hat{k})], \nonumber \\
	& {\cal H}_{DMt(2^{nd})}=-6S[(D_t^z \hat{k})\cdot (\sin(\pi/6) \hat{k})], \nonumber \\
	& {\cal H}_{DMv(1^{st})}={\cal H}_{DMv(2^{nd})}=0.
\label{H1,2HT}	
\end{align}
By imposing the above relations into Eq. (\ref{eq1}), the total energy equations are defined and the $z$-component of the intralayer DM vector is given by
\begin{equation}
	D_t^z=\frac{E_{1st}-E_{2nd}}{12 S^2 \sin(\pi/6)}.
\end{equation}
In a similar way, we can find the DM terms of model Hamiltonian for the other configurations and the interlayer DM vector of the HT bilayer is achieved by DFT-obtained total energies according to 
\begin{align}
	D_v^z&=D_{v1z}+D_{v2z}=\frac{E_{3rd}-E_{4th}}{4 S^2 \sin(\pi/6)},  \nonumber \\
	D_v^y&=D_{v1y}=\frac{E_{5th}-E_{6th}}{4 S^2 \sin(\pi/6)},  \nonumber \\
	D_v^x&=D_{v1x}+D_{v2x}=\frac{E_{8th}-E_{7th}}{4 S^2 \sin(\pi/6)}.
	\label{DMvHT}	
\end{align}
We obtain $D_t^z=1.1$ $\mu$eV, $D_v^z=3.5$ $\mu$eV , $D_v^y=-38.8$ $\mu$eV and $D_v^x=3.1$ $\mu$eV for the HT bilayer. Accordingly, the intralayer DM interaction is ignorable in comparison with the interlayer one. The results show the strongest component of the interlayer DM vector is in the $y$-direction; this means the magnetic moments of Cr atoms in the layers tend to rotate along the $x$-$z$ plane, as long as they align in the $z$-direction because of the high MAE of the bilayer.

The interlayer and intralayer DM interactions can be tuned by an external electric field, very similar to the isotropic exchange coupling. In Fig. \ref{J_DMHT} (b), it is observed that the intralayer exchange coupling is changed significantly by the electric field. Although the $z$ and $x$ components of the interlayer DM interactions are independent of the electric field, the sign and the value of $D_v^y$ vary. The sign of the $y$-component of the DM changes from negative to positive values between $2$-$2.5$ V/nm where a transition from the AFM to FM is achieved for the HT bilayer.
In fact, a phase transition needs a spin-canting between layers in the $x$-$z$ plane satisfied by the non-zero $D_v^y$ which is in the same order of the $J_v$ in the HT bilayer.
The application of the electric field shifts the atomic positions slightly, resulting in a change in the orbital couplings between atomic neighbors. Indeed, if the $e_g-t_{2g}$ coupling, the resulting FM exchange will larger than that causing the AFM exchange ($e_g-e_g$ orbital coupling), the bilayer is in the FM ground states and vice versa \cite{sivadas2018stacking, akram2021moire}. Upon increasing the electric field, an AFM to FM transition is observed in the HT bilayer CrI$_3$, showing the dominant effect of $e_g-t_{2g}$ coupling. We expect these changes to be accompanied by spin tilt, and consequently, the DM interactions representing the spin tilting ability are tuned by the electric field.

To calculate the magnetic order temperature, Curie $T_C$, or N{\'e}el $T_N$ temperature of the FM or AFM bilayer, a sample of 50$\times$50 nm$^2$ is considered and the metropolis Monte Carlo algorithms are used for the calculation of temperature-dependent magnetization (see Fig. \ref{TcHT} (a)). According to Fig. \ref{TcHT} (b), the magnetic order temperature of the HT bilayer is increased by applying the electric field that arises from the increasing of the intralayer isotropic exchange coefficient. 

The variation of $MAE=(E_\perp-E_\parallel)/\eta S^2$ as a function of the electric field is shown in Fig. \ref{TcHT} (b). It shows that the magnetic moments of the Cr atoms are in out-of-plane direction, and it is unchangeable by the external electric field.

\subsection{Exchange coefficients in the LT phase}
\begin{figure}
	\begin{center}
		\includegraphics[width=8.5cm]{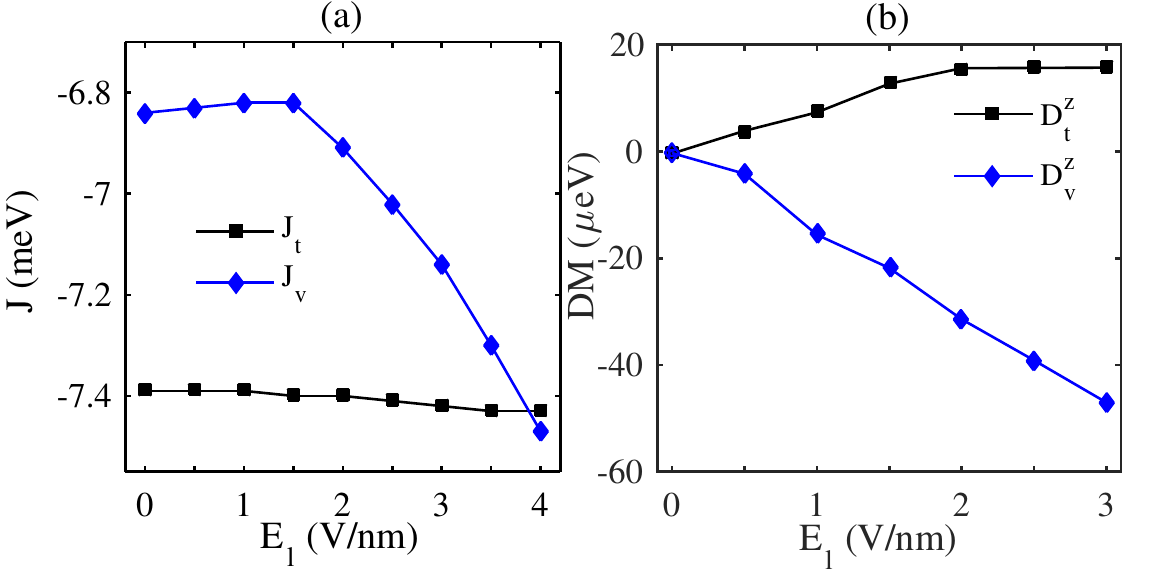}
		\caption{(Color  online) (a) The intralayer (squares) and interlayer (lozenges) isotropic exchange coefficients and (b) the $z$-component of interlayer and intralayer DM vectors of the LT bilayer CrI$_3$ as a function of the external electric field, $E_l$. The sign of the exchange coefficients remain negative by applying the electric field, while their amounts are slightly increased showing that the electric field makes the ferromagnetic coupling of layers stronger. Furthermore, the $z-$ component of the DM interaction of the LT bilayer increases by the electric field meaning that the spin-canting are in the plane of the Cr atoms. }
		\label{JDMLT}
	\end{center}
\end{figure}

Now we turn our attention to another phase of the bilayer system. Four Cr atoms in a unit cell of the LT-phase bilayer CrI$_3$ prefer to determine a parallel spin orientation in which the minimized total energy belongs to the FM magnetic configuration \cite{Morell2019}. Here, similar to the HT phase, we are interested in obtaining the symmetric and antisymmetric exchange coefficients of the LT bilayer. We should note that the intralayer nearest neighbors of the LT is equivalent to that in the HT phase, while there is only an interlayer nearest neighbor bond between the Cr atoms numbered by $2$ and $3$ in Fig. \ref{geometry} (d). In the case of the LT bilayer CrI$_3$, the first two terms of Eq. (\ref{eq1}) are extended to be as
 \begin{align}
 	{\cal H}=& \frac{1}{2}(\eta_tJ_t[({\bf S}_1\cdot{\bf S}_2)+({\bf S}_2\cdot{\bf S}_1)+({\bf S}_3\cdot{\bf S}_4)+({\bf S}_4\cdot{\bf S}_3)]+ \nonumber \\
 	&\eta_vJ_v[({\bf S}_2\cdot{\bf S}_3)+({\bf S}_3\cdot{\bf S}_2)])+\eta\gamma|S_z|^2,
 	\label{HJLT}
 \end{align}
where $\eta_t=3$ and $\eta_v=1$ are the number of the intralayer and interlayer nearest neighbors. 
We consider four different magnetic configurations of Cr atoms, as a FM (${\bf S}_1=S \hat{k}$, ${\bf S}_2=S \hat{k}$, ${\bf S}_3=S \hat{k}$, ${\bf S}_4=S \hat{k}$), 
AFM1 (${\bf S}_1=S \hat{k}$, ${\bf S}_2=S \hat{k}$, ${\bf S}_3=-S \hat{k}$, ${\bf S}_4=-S \hat{k}$), 
AFM2 (${\bf S}_1=S \hat{k}$, ${\bf S}_2=-S \hat{k}$, ${\bf S}_3=-S \hat{k}$, ${\bf S}_4=S \hat{k}$) and
AFM3 (${\bf S}_1=S \hat{k}$, ${\bf S}_2=-S \hat{k}$, ${\bf S}_3=S \hat{k}$, ${\bf S}_4=-S \hat{k}$), and thus able to find interlayer and intralayer symmetric exchange coefficients. By applying the spin vectors on Eq. (\ref{HJLT}), the relations for the total energies of the magnetic configurations are given by,
\begin{align}
	&E_{FM}=E_0+2S^2\eta_tJ_t+S^2\eta_vJ_v+\eta\gamma S^2, \nonumber \\
	&E_{AFM1}=E_0+2S^2\eta_tJ_t-S^2\eta_vJ_v+\eta\gamma S^2, \nonumber \\
	&E_{AFM2}=E_0-2S^2\eta_tJ_t+S^2\eta_vJ_v+\eta\gamma S^2, \nonumber \\
	&E_{AFM3}=E_0-2S^2\eta_tJ_t-S^2\eta_vJ_v+\eta\gamma S^2.
	\label{ELT}
\end{align}
The DFT-obtained total energies give us the symmetric exchange coefficients as
\begin{align}
	J_t&=\frac{(E_{FM}+E_{AFM1})-(E_{AFM2}+E_{AFM3})}{8S^2\eta_t}, \nonumber \\
	J_v&=\frac{(E_{FM}-E_{AFM1})}{2S^2\eta_v}.
	\label{eqJLT}
\end{align}
The results displayed in Fig. \ref{JDMLT} (a) show the interlayer and intralayer symmetric exchange couplings possess negative and consistent values with the FM ground state of the LT bilayers.
The sign of $J_t$ and $J_v$ remain negative by applying the electric field, while their amounts are slightly increased.
In fact, the electric field develops the ferromagnetic coupling of layers increase in the LT bilayer.

The third term of the Hamiltonian in Eq. (\ref{eq1}) can be written for the LT bilayer as,
\begin{align}
	{\cal H}_{DM}&=\frac{1}{2}[{\bf D}_{t12}\cdot({\bf S}_1\times{\bf S}_2)+{\bf D}_{t12'}\cdot({\bf S}_1\times{\bf S}_2')+{\bf D}_{t12''} \nonumber \\
	&\cdot({\bf S}_1\times{\bf S}_2'')+
	{\bf D}_{t21}\cdot({\bf S}_2\times{\bf S}_1)+{\bf D}_{t21'}\cdot({\bf S}_2\times{\bf S}_1') \nonumber \\
	&+{\bf D}_{t21''}\cdot({\bf S}_2\times{\bf S}_1'')
	+{\bf D}_{t34}\cdot({\bf S}_3\times{\bf S}_4)+{\bf D}_{t34'}\cdot({\bf S}_3\nonumber \\
	&\times{\bf S}_4')+{\bf D}_{t34''}\cdot({\bf S}_3\times{\bf S}_4'')+{\bf D}_{t43}\cdot({\bf S}_4\times{\bf S}_3)+{\bf D}_{t43'}\nonumber \\
	&\cdot({\bf S}_4\times{\bf S}_3')+
	{\bf D}_{t43''}\cdot({\bf S}_4\times{\bf S}_3'')+ 
	{\bf D}_{v23}\cdot({\bf S}_2\times{\bf S}_3)\nonumber \\
	&+{\bf D}_{v32}\cdot({\bf S}_3\times{\bf S}_2)
	].
	\label{HDMLT}
\end{align}
Similar to the HT phase, the symmetric of atomic positions dictates to have only the $z$-component of the intralayer DM interaction in the Hamiltonian which is given by Eq. (\ref{DMtHT}). For the interlayer DM interaction of the LT phase
\begin{equation}
	{\cal H}_{DMv}={\bf D}_{v23}\cdot({\bf S}_2\times{\bf S}_3).
\end{equation}
It should be noticed that there is a $C_3$ rotation axis on the bonding length between Cr atoms numbered by $2$ and $3$ in the unit cell of the LT bilayers (see Fig. \ref{geometry} (d)). According to the Moriya symmetry rules \cite{moriya1960anisotropic}, the DM interaction between atoms $2$ and $3$ is parallel to the $C_3$ axis which is along the $z$-direction. Therefore, we need to find the $z$-component of the interlayer DM interaction which can be obtained by two spin configurations (see Table \ref{TableLT}). Finally, the $z$-component of intralayer and interlayer DM interactions are respectively given by
\begin{align}
D_t^z=&\frac{(E_{1st}-E_{2nd})+(E_{3rd}-E_{4th})}{12S^2\sin(\pi/6)}, \nonumber \\
D_v^z=&\frac{E_{3rd}-E_{4th}}{2S^2\sin(\pi/6)}.
\end{align}

\begin{table}
	\caption{Four spin configurations for calculation of the interlayer and intralayer DM vectors of the LT bilayer CrI$_3$.}
	\begin{center}
		\begin{tabular}{|c|c|c|c|c|}
			\hline
			\hline
			Config. &  $1^{st}$ & $2^{nd}$ & $3^{rd}$ & $4^{th}$ \\
			\hline
			${\bf S}_1$ & $S\hat{i}$ & $S\hat{i}$ & $S\hat{i}$ & $S\hat{i}$ \\
			\hline
			\multirow{2}{*}{${\bf S}_2$} & $S(\cos(\frac{\pi}{6})\hat{i}+$ & $S(\cos(\frac{\pi}{6})\hat{i}-$ & $S\hat{i}$ & $S\hat{i}$ \\
			& $\sin(\frac{\pi}{6})\hat{j})$ & $\sin(\frac{\pi}{6})\hat{j})$ &   &    \\
			\hline
			\multirow{2}{*}{${\bf S}_3$} & $S\hat{i}$ & $S\hat{i}$ & $S(\cos(\frac{\pi}{6})\hat{i}+$ & $S(\cos(\frac{\pi}{6})\hat{i}-$ \\
			 &   &   & $\sin(\frac{\pi}{6})\hat{j})$ & $\sin(\frac{\pi}{6})\hat{j})$ \\
			\hline
			\multirow{2}{*}{${\bf S}_4$} & $S(\cos(\frac{\pi}{6})\hat{i}+$ & $S(\cos(\frac{\pi}{6})\hat{i}-$ & $S(\cos(\frac{\pi}{6})\hat{i}+$ & $S(\cos(\frac{\pi}{6})\hat{i}-$ \\
			&  $\sin(\frac{\pi}{6})\hat{j})$ & $\sin(\frac{\pi}{6})\hat{j})$  & $\sin(\frac{\pi}{6})\hat{j})$ & $\sin(\frac{\pi}{6})\hat{j})$ \\
			\hline
			\hline
		\end{tabular}
	\end{center}
		\label{TableLT}
\end{table}
The negligible values of $D_t^z=-0.4$ $\mu$eV and $D_v^z=-0.3$ $\mu$eV are obtained using our spin-dependent DFT-based calculations. It should be noted that there is an inversion symmetry in the center of the $2$ and $3$ atomic bond in the LT bilayer, so $D_v^z$ should be zero, and therefore, there is no any DM interaction between the layers. We investigate the effect of the external electric field on the interlayer and intralayer DM interactions of the LT phase, and the results are reported in Fig. \ref{JDMLT} (b). It is observed that the interlayer and intralayer DM interactions of the LT phase are increased by developing the electric field. It is interesting that the order of DM interaction variation is similar for both LT and HT phases, but their behavior is different. In the presence of an electric field, just the $z$-component of the DM interaction of the LT bilayer increases; this means that the spin-canting are in the plane of the Cr atoms. On the other hand, the interlayer symmetric exchange coupling, $J_v$, is much stronger than $D_v^z$, subsequently, the FM coupling survives in a competition between FM coupling of layers and in-plane canting of spins.

In the HT bilayer, the amount of the interlayer DM vector and interlayer symmetric exchange coupling are in the same order. In fact, they are in a close competition with each other and it remains in the external electric field. It is intriguing that the both of them leads to a transition from the AFM to  FM in the HT phase; the interlayer DM interaction is dominant in the $y$-direction which leads to the spin canting in the $x$-$z$-plane and, as a consequence, helps to the rotation of the spin-direction from the AFM to FM and also the sign of $J_v$ is changed by applying the electric field.

It should be noticed that the interlayer and intralayer isotropic exchange coupling in the Heisenberg Hamiltonian of the bilayer CrI$_3$ have been calculated in Ref. \cite{lei2021magnetoelectric}  which agrees well with our work. They obtained $J_t=-6.49$  meV (by DFT) and $J_v=+0.10$  meV (by experimental results from Ref. \cite{huang2017layer}) which are close to our findings ($J_t=-7.00$  meV and $J_v=+0.8$  meV) in order of magnitude and sign for HT bilayer CrI$_3$. They have also shown that the isotropic exchange couplings are tuned by the electric field. 
	There are DFT-based calculations \cite{jiang2019stacking, Morell2019, lei2021magnetoelectric}  and experimental results \cite{huang2018electrical, jiang2018controlling, Jiang2018mater} showing that the interlayer exchange coupling sign of HT phase of bilayer CrI$_3$ is changed by the electric field. One can find the DFT-obtained exchange couplings of different stacks of bilayer CrI$_3$ in the absence of an electric field in Refs. \cite{sivadas2018stacking, jiang2019stacking, leon2020strain, akram2021moire}. In contrast, the interlayer and intralayer DM interactions of the bilayer CrI$_3$  have not been calculated in previous works, and we report here their electric field tunable behavior.

\begin{figure}
	\begin{center}
		\includegraphics[width=8.5cm]{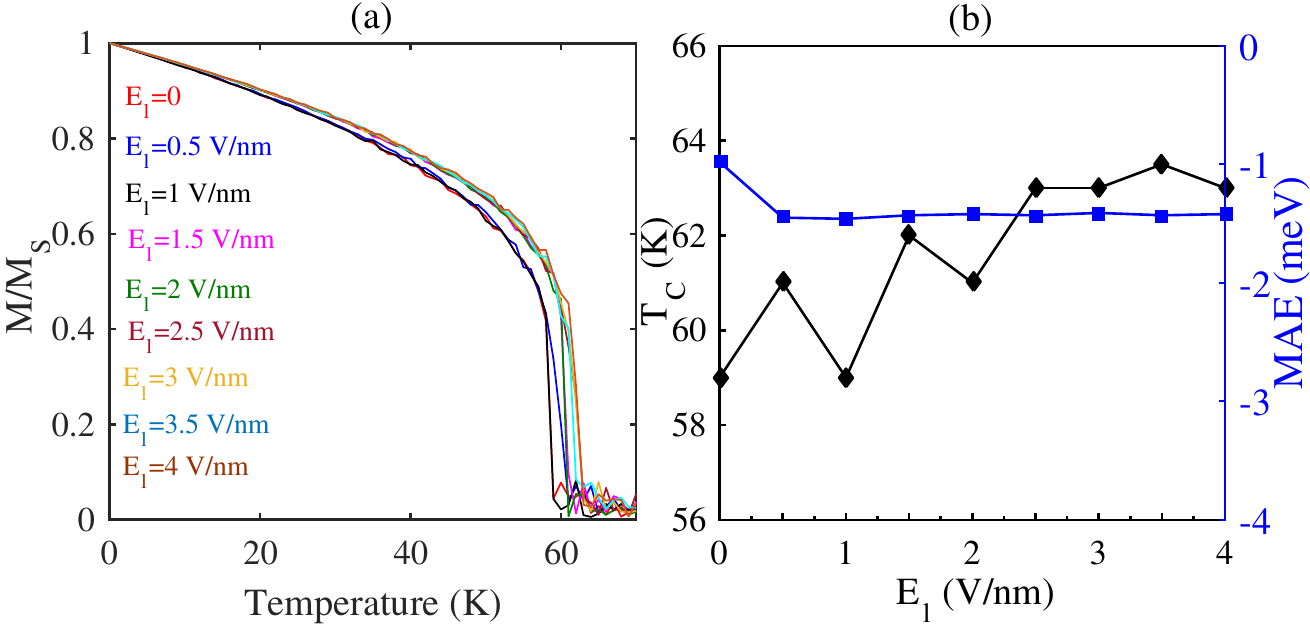}
		\caption{(Color  online) (a) The magnetization of the LT bilayers as a function of temperature under different external electric field. (b) The Curie temperature (lozenges) and magnetic anisotropy energy (squares) of LT bilayer CrI$_3$ versus external electric field. Notice that the MAE remains negative by changing the electric field, therefore, the easy axis of the LT bilayer is in the $z$-direction and its value does not change significantly.}
		\label{TcLT}
	\end{center}
\end{figure}

The temperature dependence of the magnetic moments of the Cr atoms, $M(T)$ in the LT bilayer is illustrated in Fig. \ref{TcLT} (a). It is shown that $M(T)$ is nearly independent of the electric field. The variation of the Currie temperature, $T_C$, by the electric field arises directly from the variation of $J_t$ and it is gradually increased in a more intense electric field (see Fig. \ref{TcLT} (b)). The MAE of the LT phase as a function of the electric field is shown in Fig. \ref{TcLT} (b). The MAE remains negative by varying the electric field so the easy axis of the LT bilayer lies in the $z$-direction and its value does not change significantly.

We obtained a N{\'e}el temperature of 44 K for the AFM-HT bilayer and a Curie temperature of 59 K for the FM-LT bilayer CrI$_3$, which are tuned by the electric field due to the variation in the parameters of the spin model Hamiltonian. Increasing the number of layers from monolayer to bulk increases the critical temperature between 45 K and 61 K \cite{huang2017layer, kashin2020orbitally}. A N{\'e}el temperature of 45 K was reported experimentally for the HT bilayer \cite{huang2018electrical}. The larger Curie temperature is due to the stronger FM interlayer exchange coupling in LT bilayer CrI$_3$ \cite{wang2021systematic}.

Although the magnetic coupling coefficients of both the LT and HT bilayers are tuned, the magnetic ground state of the HT bilayer can be manipulated by the electric field. To perceive the possibility of manipulating magnetic textures in the bilayer CrI$_3$, we simulate their spin dynamics in the presence of the external electric and magnetic fields and the emergence of the magnetic domains and skyrmion patterns are explored.

To obtain the spin dynamics of the bilayer, the atomic position of the system is constant while different atomic magnetic moments are appeared by time evolutions \cite{evans2014atomistic}. Here, we obtained that the formation energy is 1000 times more than the exchange energy in the bilayer CrI$_3$ which guarantees the stability of the atomic structure under spin evolutions. In fact, we want to mention that the spin evolution happens in a constant atomic structure in any electric field. Under the electric field, the atomic positions are relaxed by DFT-based calculations and leading variation of the interband and intraband isotropic exchange and DM coefficients, so the spin dynamics are calculated in the new structure.

\subsection{Spin dynamics of the HT phase}
\begin{figure}
	\begin{center}
		\includegraphics[width=8.5cm]{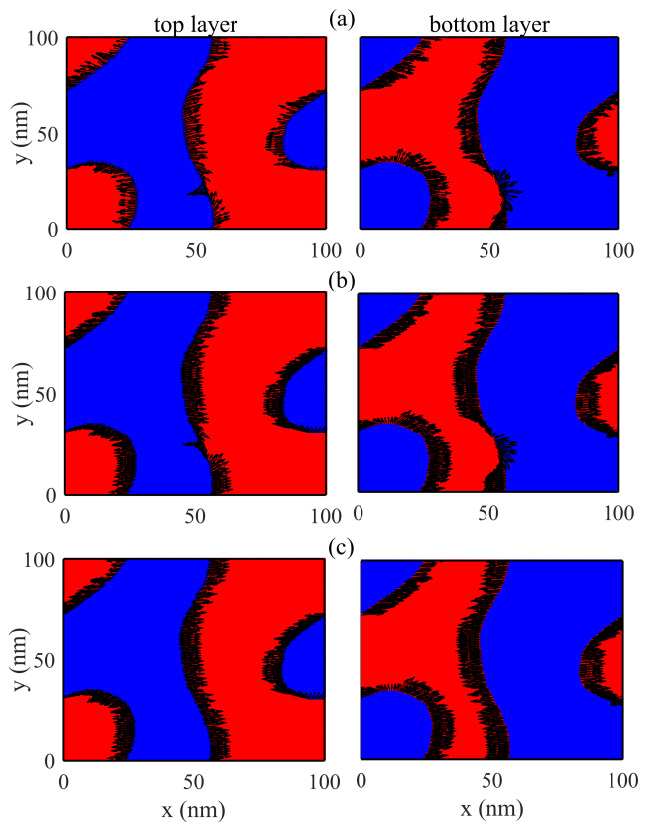}
		\caption{(Color  online) The magnetic texture of the top and bottom layers of the HT bilayer in $E_l=0$ V/nm, $B= 0$, (a) $T= 0.3$ K, ${\text {time}}=2$ ns, (b) $T= 0$, ${\text {time}}=2.2$ ns and (c) $T=0$, ${\text {time}}=3$ ns.
		The blue (red) areas show where the magnetic moments of Cr atoms are in the -z(+z)-direction. ${M}_{\perp}$ is represented by black arrows on the domain walls.
			 The created magnetic texture is stable in size, but ${M}_{\perp}$ is slowly changed to be perpendicular to the domain wall by the time evolution.}
		\label{3tE0H0Ht}
	\end{center}
\end{figure}

In the HT bilayer, the AFM coupling between layers is dictated in the spin dynamics of the system. The spin dynamics of the top and bottom layers are plotted in Fig. \ref{3tE0H0Ht}. In $T=0.3$ K, it is observed that the bilayer is divided into two areas in which layers are coupled in form of an anti-ferromagnetic. While the bottom layer has a magnetic moment in the $+z$-direction, the top layer has a magnetic moment in the $-z$-direction and vice versa; this means that the AFM order is constant between the layers. This regularity is also established on the domain walls, where the perpendicular components of the magnetic moments, $M_\perp$, of two layers are in opposite directions. The results of the simulations show that the created magnetic texture is stable in size, but $M_\perp$ is slowly changed to be perpendicular to the domain wall by the time evolution as shown in Figs. \ref{3tE0H0Ht} (a)-(c). Thus, an ultrathin N{\'e}el-type domain wall appears on both the top and bottom layers. It is clear that the zero magnetic moments are obtained for the HT bilayer, but there is an interesting microscopic magnetic pattern on each layer.

\begin{figure}
	\begin{center}
		\includegraphics[width=8.5cm]{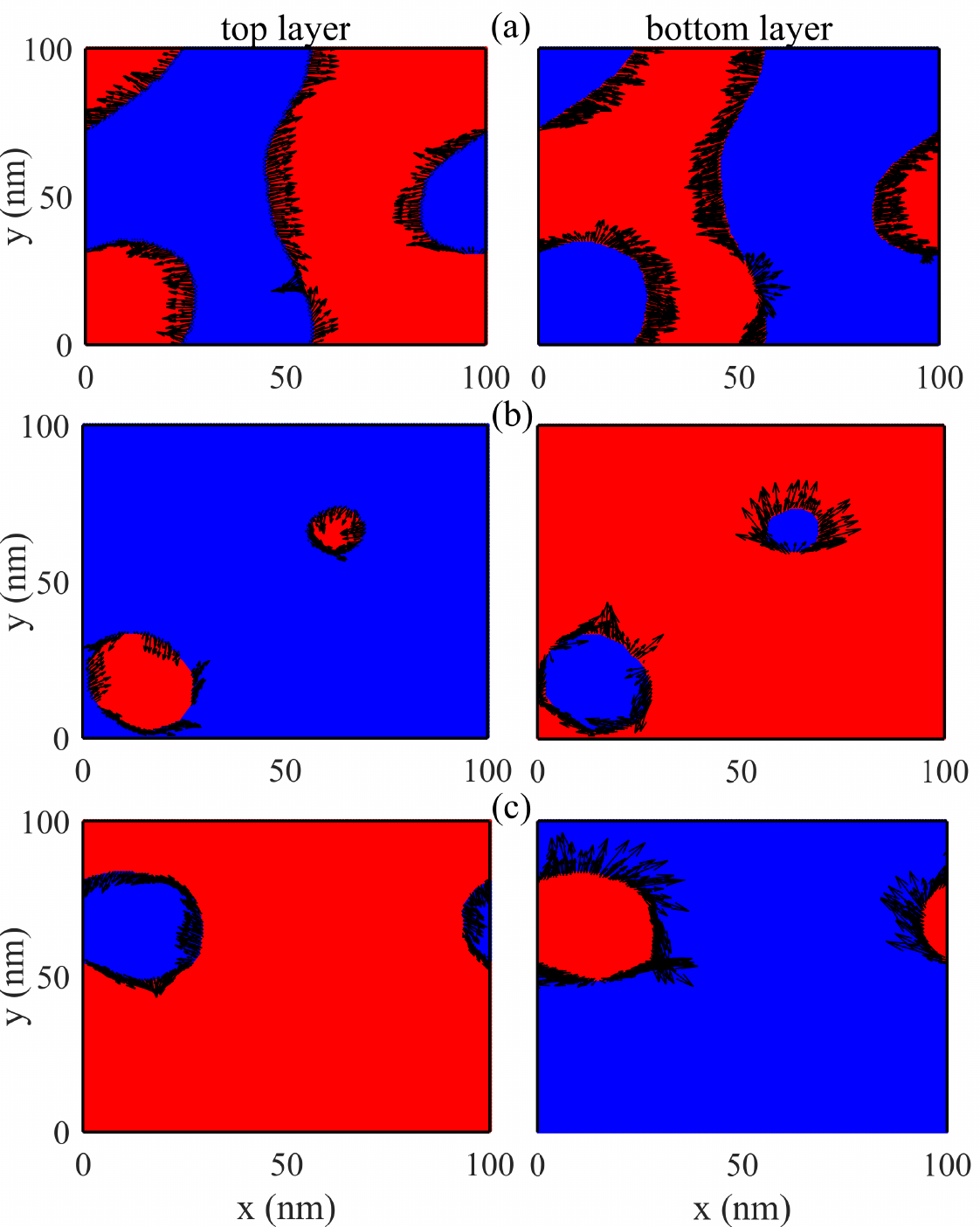}
		\caption{(Color  online) The magnetic texture of the top and bottom layers of the HT bilayer in (a) $E_l=0$, (b) $E_l=1$ V/nm and (c) $E_l=2$ V/nm, $T= 0.3$ K, $B= 0$ and ${\text {time}}=2$ ns. The skyrmion patterns occur around the magnetic moments, while their chirality is different for the top and bottom layers. The places and sizes of magnetic domains are the same in the top and bottom layers, however, the direction of the magnetic moments are exactly opposite for inside and outside of the domain walls.}
		\label{3EH0Ht}
	\end{center}
\end{figure}

We obtain the symmetric and asymmetric exchange coupling coefficients of the spin Hamiltonian depending on the external electric field, therefore, we explore the variation of the spin dynamics of the bilayer by the electric field shown in Figs. \ref{3EH0Ht} (a)-(c). In $E_l=1$ V/nm, the magnetic order of the HT bilayer is the AFM and $J_t$ and $J_v$ are near to their values in the absence of the electric field case, while the value of $D_t^z$ is increased to $25.5$ $\mu$eV, therefore the new spin texture is created (see Fig. \ref{3EH0Ht} (b)). If we focus on one of the layers, we see that the wider magnetic domains in $E_l=0$ case (Fig. \ref{3EH0Ht} (a)) changes to two smaller quasi-circle magnetic domains with finite diameters (nearly $25$ nm and $10$ nm). Skyrmion patterns occur around the magnetic moments, while their chirality is different for the top and bottom layers. The places and sizes of magnetic domains are the same in the top and bottom layers, however, the direction of the magnetic moments are exactly opposite for inside and outside of the domain walls. We obtain these stable quasi-circle magnetic domains under the time evolution.

For the HT bilayer, the amount of the intralayer DM interaction reaches to $41$ $\mu$eV by increasing the electric field to $E_l=2$ V/nm, and the interlayer isotropic exchange coupling is equal to $-20$ $\mu$eV. Although the value of $J_v$ is reduced, the negative sign shows the bilayer is still in the AFM phase. These changes lead to a new spin texture in the bilayer. In Fig. \ref{3EH0Ht} (c), 30 nm-diameter quasi-circle magnetic domains are created in the both top and bottom layers with opposite spin orientation, so the AFM configuration of the bilayer is conserved. Most importantly, the diameter, position, number of the magnetic domain and the domain wall chirality are related to the external electric field. In fact, the magnetic texture of the HT bilayer can be manipulated by an external electric field owing to the variation of the DM and isotropic exchange coupling interactions. Therefore, stable magnetic domains are constituted in the HT bilayer that possess fascinating microscopic information, while the total magnetism is zero in the macroscopic level. 

\begin{figure}
	\begin{center}
		\includegraphics[width=8.5cm]{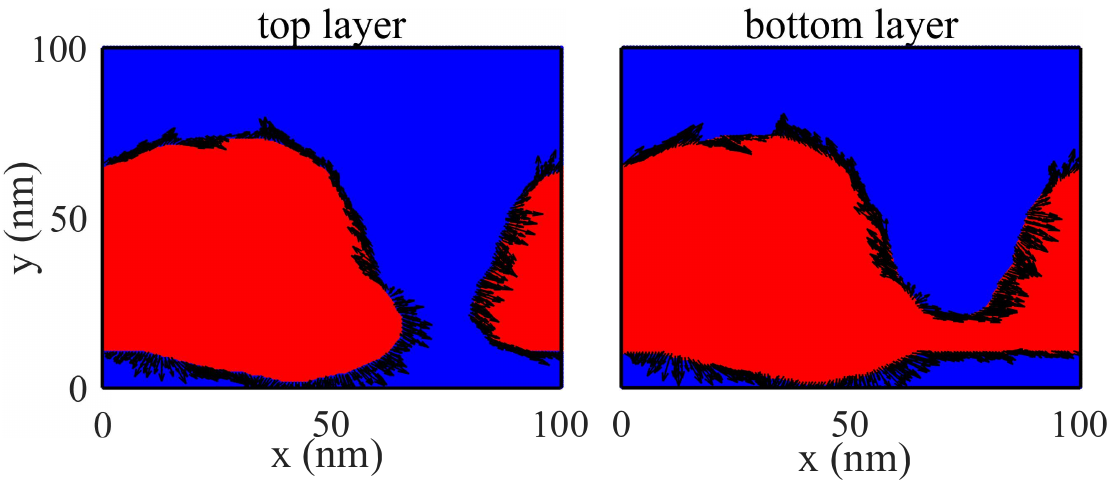}
		\caption{(Color  online) The magnetic texture of the top and bottom layers of HT bilayer in $E_l=3$ V/nm, $T= 0.3$ K, $B= 0$ and ${\text {time}}=2$ ns. Notice that the the magnetic domain in the FM bilayer is stable under the time evolution.}
		\label{E3H0Ht}
	\end{center}
\end{figure}

In $E_l=3$ V/nm, the layers are coupled in a form of a ferromagnetic. This phase transition is obviously observed in the spin dynamics simulation which both layers have the same spin orientation (see Fig. \ref{E3H0Ht}) and originates from the positive sign of  $J_v$. On the other hand, the perpendicular-component of the magnetic moments of the top and bottom layers are similar on the domain walls. The spin dynamics simulations show that the obtained magnetic domain in the FM bilayer is still stable under the time evolution.
\begin{figure}
	\begin{center}
		\includegraphics[width=8.5cm]{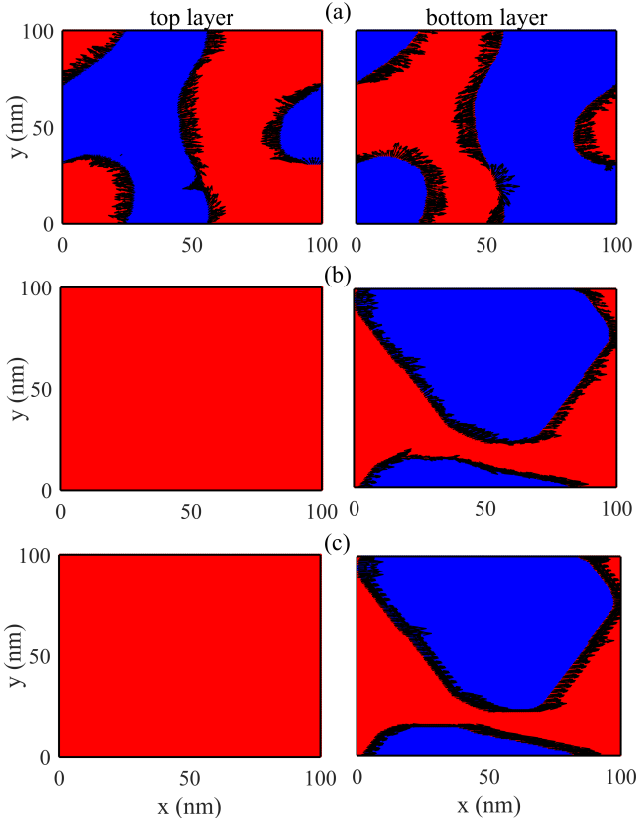}
		\caption{(Color  online) The spin dynamic results of the HT bilayer in $E_l=0$ V/nm, (a) $B=0$,  $T= 0.3$ K and $time=2$ ns,  (b) $B= 0.5$ T,  $T= 0.3$ K and ${\text {time}}=2$ ns and (c) $B= 0.5$ T,  $T= 0$ K and ${\text {time}}=3$ ns. the external magnetic field is in competition with the interlayer exchange coupling and desire to bring the spin of the Cr atoms parallel to the external magnetic field. }
		\label{E0H500Ht}
	\end{center}
\end{figure}

According to Figs. \ref{E0H500Ht} (a)-(c), our simulation results show that the external magnetic field changes the spin dynamic of the bilayer. In the absence of an external electric field, Fig.\ref{E0H500Ht} (b) shows that there is an area with FM configurations in the AFM background of the bilayer. Indeed, the external magnetic field is in competition with the interlayer exchange coupling and desire to accompany the spin of the Cr atoms parallel to the external magnetic field. It is not completely winner in the competition when $B_{ext}$ is less than $0.5$ T. By increasing the external magnetic field to $B_{ext}=2$ T, the layers gain parallel magnetic moments in the direction of the applied magnetic field.
\begin{figure}
	\begin{center}
		\includegraphics[width=8.5cm]{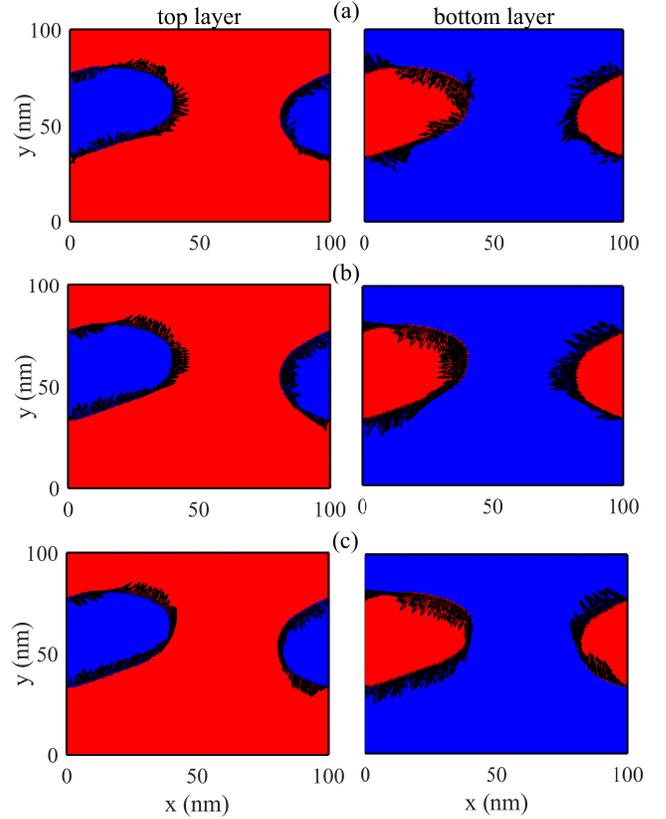}
		\caption{(Color  online) The spin dynamic results of the HT bilayer in $E_l=2$ V/nm, $B=0.01$ T  (a)  $T= 0.3$ K, $time=2$ ns,  (b) $T= 0$ K, ${\text {time}}=2.2$ ns and (c) $T= 0$ K, ${\text {time}}=3$ ns. Notice, here, the applied magnetic field can not create a phase transition in the magnetic order of the HT bilayer. }
		\label{E2H100H3t}
	\end{center}
\end{figure}

Another example is the system when exposed to an external electric field, $E_l=2$ V/nm. In $B_{ext}=0.5$ T, the AFM pattern is completely disappeared and the magnetic moments of the Cr atoms are parallel. In the competition between exchange coupling coefficients and the Zeeman energy, the Zeeman energy of the strong magnetic field has conquered. In the presence of $0.01$ T external magnetic field, the simulation results (Figs. \ref{E2H100H3t}(a)-(c)) show that the quasi-circular magnetic domain form converts to a quasi-elliptical shape, but the AFM configuration between the top and bottom layers is conserved because the applied magnetic field is not large enough to create a phase transition in the magnetic ordering of the bilayer. Interestingly, the obtained magnetic domain and almost domain wall do not change by the time evolution after cooling time (2 ns). Accordingly, a stable magnetic texture can be obtained and tuned by a weak external magnetic field, while the strong magnetic field leads to the magnetic phase transition in the bilayer.
\subsection{Spin dynamic of the LT phase}
\begin{figure}
	\begin{center}
		\includegraphics[width=8.5cm]{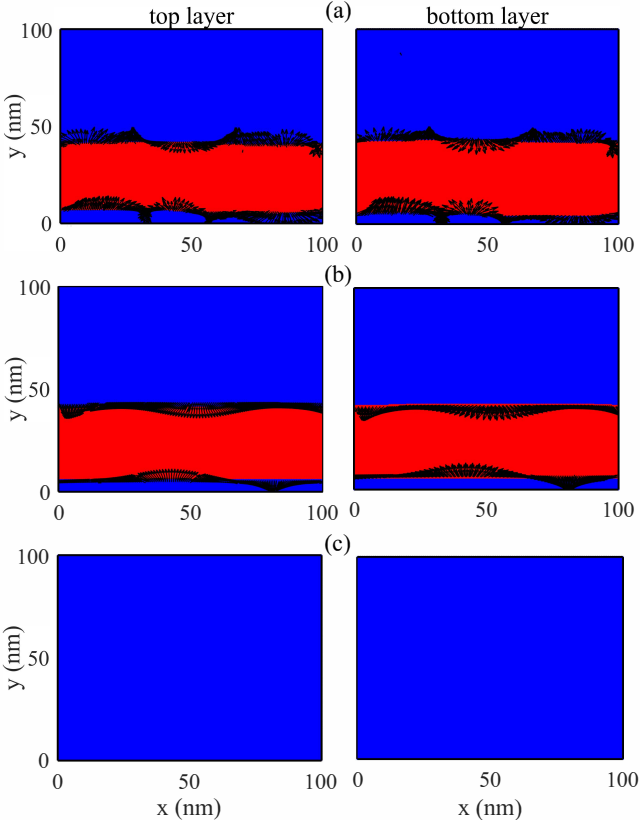}
		\caption{(Color  online) The spin dynamic results of the LT bilayer in $E_l=0$ V/nm, $B_{ext}=0$ T  (a)  $T= 0.3$ K and $time=2$ ns,  (b) $T= 0$ K and ${\text {time}}=3$ ns and (c)   $T= 0.3$ K and ${\text {time}}=2$ ns in the absence of a demagnetization field. A 1D metastable spin-wave is created at the boundary and the chirality of the spin wave is changed smoothly in time. In this case, the Cr atoms have parallel magnetic moments due to strong interlayer and intralayer isotropic exchange couplings.}
		\label{3tE0H0LT}
	\end{center}
\end{figure}

In the LT bilayer, a strong negative interlayer exchange coupling leads to the creation of the FM magnetic configuration. The spin dynamic simulation, in Fig. \ref{3tE0H0LT} (a) shows two wide areas with opposite spin directions in each layer while the magnetic moments of the layers are actually analogous. In the boundary of the two areas, the perpendicular magnetic moments of the Cr atoms are dominated and their chirality is the same in both layers. In fact, a 1D spin-wave is created at the boundary and the results show that the chirality of the spin-wave is changed smoothly in time (compare Fig. \ref{3tE0H0LT} (a) and (b)). On the other hand, the created areas sizes are stable by the time evolution, so we can consider that the 1D in-plane spin-wave is metastable (due to the change of chirality). In the absence of an electric field, the ignorable DM interaction cannot create a magnetic domain, and the observed spin-wave is due to the demagnetization field. For clarity, we report the results of the spin dynamic simulation in the absence of a demagnetization field in Fig. \ref{3tE0H0LT} (c). It is obviously observed that the magnetic domain is disappeared and all the Cr atoms of both layers have parallel magnetic moments due to strong interlayer and intralayer isotropic exchange couplings of the LT bilayer CrI$_3$.

\begin{figure}
	\begin{center}
		\includegraphics[width=8.5cm]{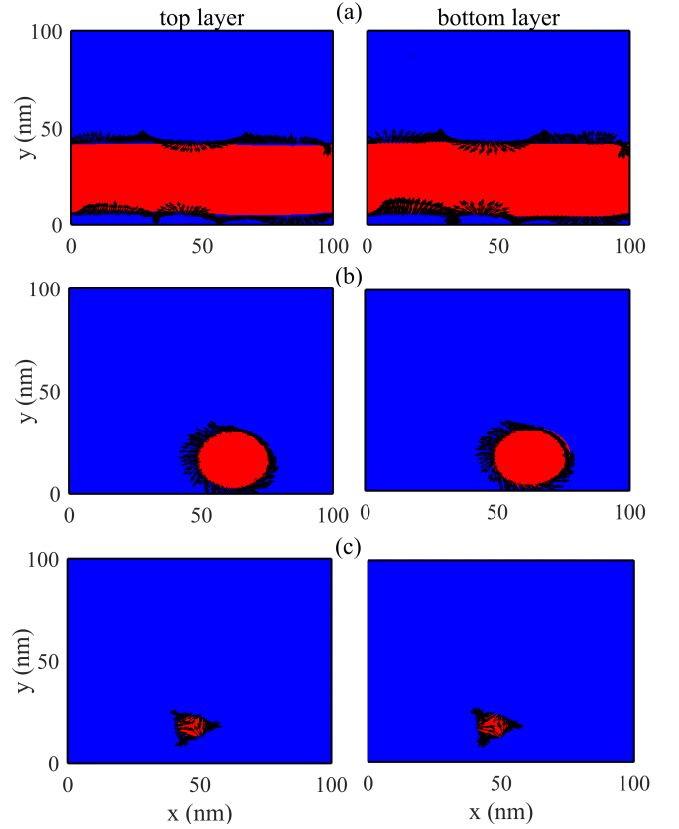}
		\caption{(Color  online) The magnetic texture of the top and bottom layers of the LT bilayer in (a) $E_l=0$, (b) $E_l=1$ V/nm and (c) $E_l=2$ V/nm, $T= 0.3$ K, $B= 0$ and ${\text {time}}=2$ ns. The existence of a similar quasi-circle magnetic domain in the top and bottom layers in the presence of the electric field are seen and the meta-stable magnetic domain is disturbed by increasing the DM interaction.}
		\label{3EH0LT}
	\end{center}
\end{figure}

\begin{figure*}
	\begin{center}
		\includegraphics[width=17cm]{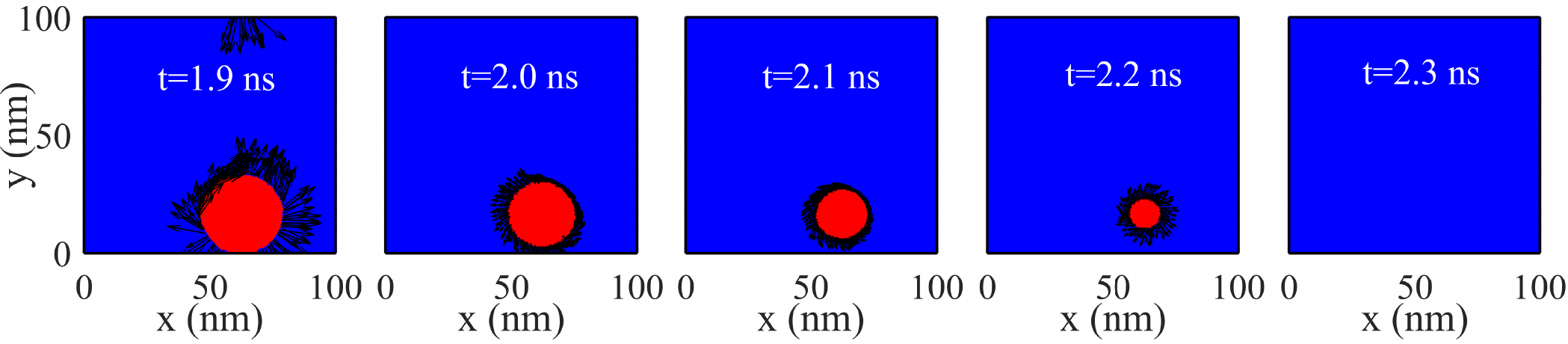}
		\caption{(Color  online) The time evolution of the top layer of the LT phase under $E_l=1$ V/nm, $T= 0$ K and $B= 0$. The bottom layer has a similar spin dynamic results due to the FM coupling of the layers.}
		\label{timeH0LT}
	\end{center}
\end{figure*}

The interlayer and intralayer DM interactions are increased by applying the external electric field, while the sign of the exchange coupling and their coefficients values remain practically constant in the LT bilayer CrI$_3$. The spin dynamic simulations show the existence of a similar quasi-circular magnetic domain shape in the top and bottom layers in the presence of the electric field. Hence, the meta-stable magnetic domain is disturbed by increasing the DM interaction in $E_l=1$ V/nm and $E_l=2$ V/nm, see Figs. \ref{3EH0LT} (a)-(c). The results show that this quasi-circular magnetic domains are disappeared by cooling the system and the magnetic moments of the Cr atoms gained parallelized, see Fig. \ref{timeH0LT} for the LT bilayer under $E_l=1$ V/nm. We can assume that the DM interactions leads to the creation of a quasi-circular domain shape in the layers, but its value is extremely smaller than the exchange coupling to stabilize it in the system, therefore, the obtained magnetic domain is unstable and disappears rapidly in the LT bilayer.

The spin dynamics in the twisted bilayer CrI$_3$ was recently calculated using an arbitrary DM interaction and a local moir{\'e} field instead of an isotropic exchange coupling between the layers \cite{akram2021moire}  while ignoring the demagnetization field. They showed that the twisted texture depends on the twist angle. It is worth noting that applying an electric field is easier than twisting the double layer CrI$_3$ in experiments. It is experimentally shown that the moir{\'e} magnetic domain in AFM twisted bilayer magnetic crystals can be controlled by gate voltages \cite{xu2021emergence}. Despite a recent interest in studying spin dynamics in 2D magnets \cite{ akram2021moire, xu2021emergence, wahab2021quantum, tong2018skyrmions, skyrmion2, Behera, abdul2021domain}  and their promising application in new generation magnetic domain-based devices \cite{ allwood2005magnetic, velez2019high, luo2020current}, there is no report of spin texture formation in the bilayer CrI$_3$ by considering all DFT-obtained isotropic and anisotropic exchange couplings in the presence of an electric field. Here we calculate the spin dynamics of a large bilayer CrI$_3$  as an experimental domain by solving the LLG equations where the demagnetizing field is important to find the best magnetic ground state of the system. The results show that the magnetic domain in the HT bilayer CrI$_3$ can be tuned by the electric field, which is consistent with the experimental results \cite{xu2021emergence}.

\section{Conclusion}
The interlayer and intralayer isotropic exchange coupling coefficients and DM interactions are obtained by DFT-based calculations for bilayer CrI$_3$.
We have shown that in the presence of an external electric field, interlayer DM interactions have a critical impact on the magnetic phase transition of the HT bilayer.
The effects of the external electric field on the parameters of the spin-model Hamiltonian and temperature-dependent magnetic moments of bilayers are explored. Most importantly, a stable magnetic domain in the HT bilayer CrI$_3$ which can be manipulated by the electric and magnetic fields can be constituted owing to the tunable interlayer and intralayer exchange coupling and DM interactions. The spin dynamics simulations show that the HT bilayer CrI$_3$ can represent a promising candidate for spintronic and logic memory applications.
The interlayer exchange coupling, on the other hand, is only weakly affected by the external electric field, hence there is no stable magnetic domain in the LT bilayer CrI$_3$. Current experiments can investigate these findings. Furthermore, when disorder and magnetic impurity are taken into account, our results can be extended in a system.  
\begin{acknowledgments}
We thank Z. Torbatian and S. Heydari for fruitful discussions. R. A thanks for support from the Australian Research Council Centre of Excellence in Future Low-Energy Electronics Technologies (project number CE170100039).
\end{acknowledgments}

\nocite{apsrev41Control}
\bibliographystyle{apsrev4-1}
\bibliography{saha}

\end{document}